\documentclass[twocolumn]{aastex631}

\newcommand\tnote[1]{$^{\rm #1}$}

\newcommand{\jeff}[1]{\textcolor{red}{[JH]}}
\newcommand{\kam}[1]{\textcolor{red}{[KAM]}}

\usepackage{amsmath}
\usepackage{booktabs}
\usepackage{multirow}
\usepackage{threeparttable}
\usepackage{ulem}
\usepackage{soul}
\usepackage{comment}

\begin{document}

\title{Evolution of the Termination Region of the Parsec-Scale Jet of 3C~84 Over the Past 20 Years}

\correspondingauthor{Sascha Trippe}
\email{trippe@snu.ac.kr}
\email{mckqm2001@gmail.com, jhodgson@sejong.ac.kr}

\author[0000-0001-9799-765X]{Minchul Kam}
\affiliation{\normalfont Department of Physics and Astronomy, Seoul National University, Gwanak-gu, Seoul 08826, Republic of Korea}

\author[0000-0001-6094-9291]{Jeffrey A. Hodgson}
\affiliation{\normalfont Department of Physics and Astronomy, Sejong University, 209 Neungdong-ro, Gwangjin-gu, Seoul, Republic of Korea}
\affiliation{\normalfont Korea Astronomy and Space Science Institute, Daedeokdae-ro 776 Daejeon, Republic of Korea }

\author[0000-0001-6558-9053]{Jongho Park}
\affiliation{\normalfont Department of Astronomy and Space Science, Kyung Hee University, 1732, Deogyeong-daero, Giheung-gu, Yongin-si, Gyeonggi-do 17104, Republic of Korea
}
\affiliation{\normalfont Korea Astronomy and Space Science Institute, Daedeok-daero 776, Yuseong-gu, Daejeon 34055, Republic of Korea}
\affiliation{\normalfont Institute of Astronomy and Astrophysics, Academia Sinica, P.O. Box 23-141, Taipei 10617, Taiwan}

\author[0000-0002-2709-7338]{Motoki Kino}
\affiliation{\normalfont Kogakuin University of Technology $\&$ Engineering, Academic Support Center, 2665-1 Nakano, Hachioji, Tokyo 192-0015, Japan}
\affiliation{\normalfont National Astronomical Observatory of Japan, 2-21-1 Osawa, Mitaka, Tokyo 181-8588, Japan}

\author[0000-0003-0292-3645]{Hiroshi Nagai}
\affiliation{\normalfont National Astronomical Observatory of Japan, 2-21-1 Osawa, Mitaka, Tokyo 181-8588, Japan}
\affiliation{\normalfont The Graduate University for Advanced Studies, SOKENDAI, 2-21- Osawa, Mitaka, Tokyo 181-8588, Japan}

\author[0000-0003-0465-1559]{Sascha Trippe}
\affiliation{\normalfont Department of Physics and Astronomy, Seoul National University, Gwanak-gu, Seoul 08826, Republic of Korea}
\affiliation{\normalfont SNU Astronomy Research Center, Seoul National University, Gwanak-gu, Seoul 08826, Korea}

\author[0000-0002-5104-6434]{Alexander Y. Wagner}
\affiliation{\normalfont Center for Computational Science, Tsukuba University, 1-1-1 Tennodai, Tsukuba, Ibaraki 305-8577, Japan}

\begin{abstract}
\noindent We present the kinematics of the parsec-scale jet in 3C~84 from 2003 November to 2022 June observed with the Very Long Baseline Array (VLBA) at 43 GHz. We find that the C3 component, a bright feature at the termination region of the jet
component ejected from the core in 2003, has maintained a nearly constant apparent velocity of $0.259 \pm 0.003c$ over the period covered by observations. We observe the emergence of four new subcomponents from C3, each exhibiting apparent speeds higher than that of C3. Notably, the last two subcomponents exhibit apparent superluminal motion, with the fastest component showing an apparent speed of $1.22 \pm 0.14c$. Our analysis suggests that a change in viewing angle alone cannot account for the fast apparent speeds of the new subcomponents, indicating that they are intrinsically faster than C3. We identify jet precession (or reorientation), a jet--cloud collision, and magnetic reconnection as possible physical mechanisms responsible for the ejection of the new subcomponents. 
\end{abstract}

\keywords{galaxies: active --- galaxies: jets --- radio continuum: galaxies}

\section{Introduction} \label{sec:intro}

3C~84 is a nearby radio AGN located at the center of the elliptical galaxy NGC~1275 \citep[z = 0.0176;][]{strauss92}. It exhibits multiple radio lobes on both sides at parsec scales as the result of intermittent jet activity \citep[e.g.,][]{pedlar90, wrb94, nagai10, fujita17}. A long-lived jet component emerged with a radio outburst in 2003 November \citep{nagai10, suzuki12}. This is a rare case in which the restart activity of AGN jets was observed. This provides a unique opportunity to investigate how the jet, and in particular its termination region, evolves on (sub-)parsec scales right after the jet is formed. A bright feature at the jet termination region, referred to as C3, exhibits a bowshock-like structure containing a hotspot. The hotspot is considered to be a region containing a strong shock  due to the interaction between the jet and the surrounding dense ambient medium, where particles are efficiently accelerated \citep{nagai16}. This suggests that C3 moves through a dense surrounding medium. As a result, C3 is not accelerated to apparent superluminal speeds which are seen in some classes of AGN jets such as blazars. Instead, it maintains a mildly relativistic speed with $\beta_{\rm app}\sim0.2-0.3$ \citep{nagai10, suzuki12, hiura18, hodgson18, weaver22}. This is consistent with the typical speeds of hotspots, $\beta_{\rm app}=0.04-0.4$, in other young radio sources \citep{odea21}. 

In this paper, we present the discovery of subcomponents ejected at C3 with superluminal speeds. We also analyze the kinematics of C3, covering its entire 20-year time span. In Section \ref{sec:Data}, we summarize the Very Long Baseline Array (VLBA) data used for the analyses. In Section \ref{sec:analysis}, we describe the methods used for the analysis and present the kinematic properties of C3 and new subcomponents. In Section \ref{sec:Discussions}, we interpret the apparent motions of C3 and the new subcomponents, and discuss the mechanisms for the ejection of the subcomponents. A summary of our findings and conclusions follow in Section \ref{sec:Conclusions}. 

Given that 3C~84 is located in the local universe, we adopt cosmological parameters $H_{\rm 0}= \rm 73.30\ km\ s^{-1}\ Mpc^{-1}$ and $q_{0}=-0.51\pm0.024$ obtained from the Type Ia supernovae in the nearby universe \citep{riess22}, which correspond to $\Omega_{\rm m}=\rm 0.327$ and $\Omega_{\rm \Lambda}=\rm 0.673$ in a flat universe. This leads to a linear scale of 1 mas = 0.34 pc at a distance of 73 Mpc. It is noteworthy that the distance of 73 Mpc is consistent with the distance derived from the variability (light-crossing) timescale of C3, $\rm D=72^{+5}_{-6}\ Mpc$ \citep{hodgson20}.

\begin{figure*}[ht!]
\centering 
\includegraphics[width=\textwidth]{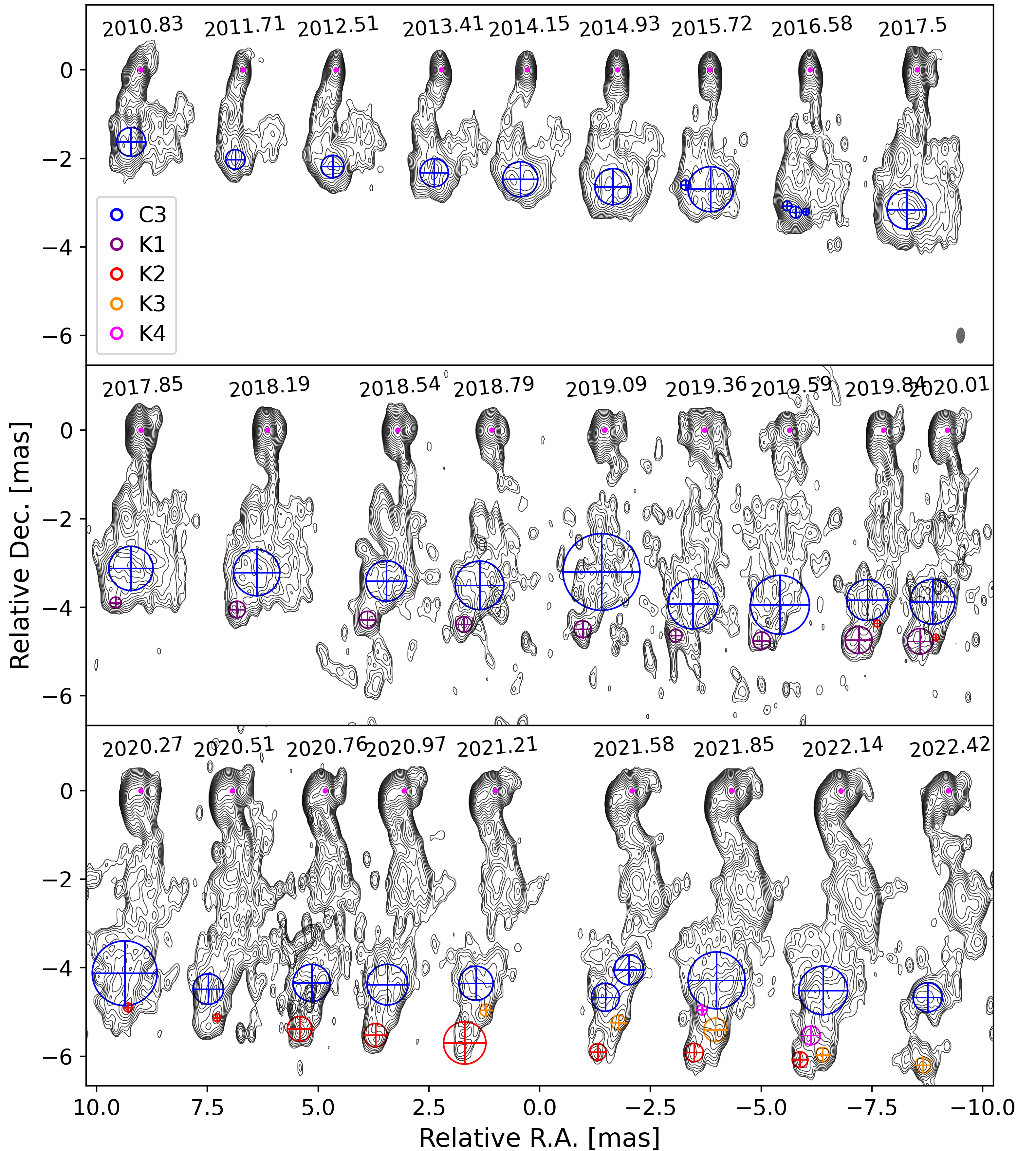}
\caption{Selected CLEAN images of 3C~84. The contours start from 12 times the rms noise in each epoch and increase by a factor of $\sqrt{2}$. The MODELFIT components are overlaid as circular Gaussian components with plus signs onto the clean images. While MODELFITs were performed to the entire jet, only the relevant components to the C3 region are shown. C3, K1, K2, K3, and K4 are colored blue, purple, red, orange, and magenta, respectively. The magenta colored points are the reference points used to determine the positions of C3 and K1-K4.
The observation dates, given in decimal years, are shown above each image. The top row features selected images obtained between 2010.83 and 2017.50. During this period, C3 is the farthest jet component from the jet base. Horizontal spacing is proportional to elapsed time. In the center and bottom rows, selected images obtained between 2017.85 and 2022.42 are shown. During this period, new subcomponents emerge from C3. The horizontal spacing is proportional to the elapsed time albeit on a scale which is different from the top row. 
\label{fig:series}}
\end{figure*}

\section{Observations} \label{sec:Data}

We analyzed the VLBA 43~GHz archival data\footnote{\url{https://www.bu.edu/blazars/BEAM-ME.html}} (hereafter BU data) obtained as part of the VLBA-BU-BLAZAR program \citep{jorstad16}. 3C~84 was observed along with 32 other sources nearly every month since 2010 November \citep{weaver22}. 3C 84 was observed for 20--30 minutes per integration (7--10 scans of $\sim$3 minutes duration each), spread over an 7--8 hour time span to obtain a good (u,v) coverage. The data was typically recorded with a single 20--64 MHz intermediate frequency (IF) band from 2010 November to 2014 May. Subsequently, it was recorded with four 64 MHz IFs from 2014 June to 2020 July, and with four 122--124 MHz IFs from 2020 August to 2022 June. We applied uniform weighting, leading to the average beam size being 0.32$\times$0.15 mas with a position angle (P.A.) of 2.5 degrees and an average rms noise level of $\sim$3 mJy. We used all 112 epochs obtained between 2010 November and 2022 June. This high cadence of observations over more than a decade enabled us to investigate the kinematics of 3C 84 in unprecedented detail. 
In addition, we obtained earlier data between 2003 November to 2008 November from \citet{suzuki12}. Since these data were also obtained from the VLBA 43 GHz observations, they can be merged into the results obtained from the BU data in a straightforward manner. 

While the (u,v) coverage of the BU data is well-suited for kinematics, it is not sufficient to reliably obtain the flux densities of the complex structure of 3C~84. Consequently, we focus on the kinematics in this paper.

\begin{figure*}[h!]
\centering 
\includegraphics[width=\textwidth]{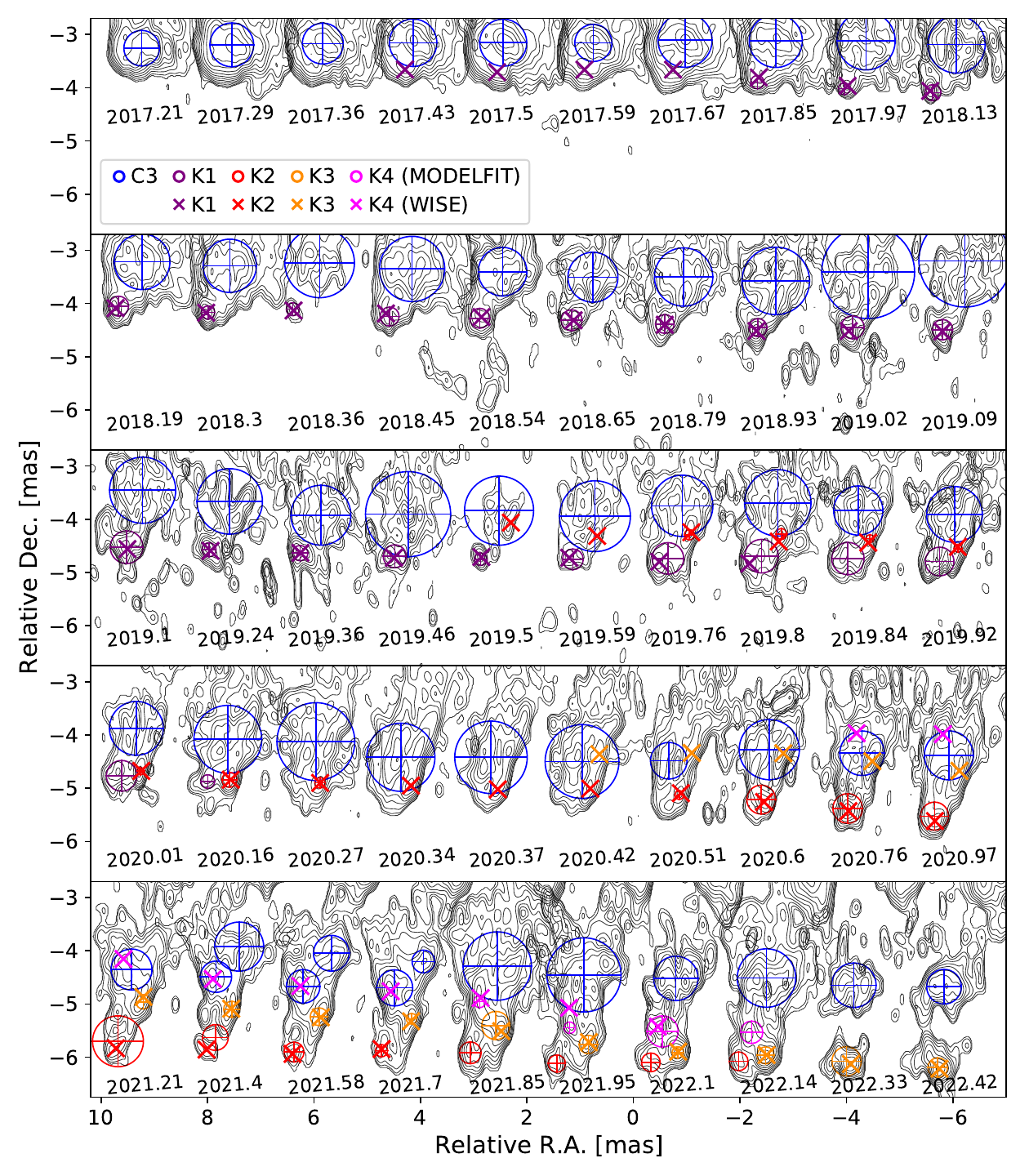}
\caption{Enlarged CLEAN images of the jet termination region of 3C~84 obtained from 2017.21} to 2022.42. The contour levels are the same as in Figure 1. SSPs identified with WISE at a scale of 0.16 mas are marked by crosses, the MODELFIT components are shown as ellipses. C3, K1, K2, K3, and K4 are colored blue, purple, red, orange, and magenta, respectively. The observation dates, in decimal years, are shown at each image. 
\label{fig:overplot_enlarged}
\end{figure*}

\section{Kinematic Analysis and Results} \label{sec:analysis}

The task MODELFIT in Difmap is widely used to locate jet components in VLBI images, allowing us to infer their motions from comparison of multiple epochs \citep{shepherd97}. However, the jet termination region of 3C~84 is well known for its complex structure. Moreover, the new subcomponents ejected from C3 have small sizes and flux densities. Consequently, it is essential to cross-identify them with different methods for reliable identification. A package called the wavelet-based image segmentation and evaluation (WISE) has been developed to track the displacements of such complex structure of jets \citep{mertens15}. Here we describe how we used both methods to reliably track the jet termination region of 3C~84. \\

\subsection{MODELFIT} 
We used a set of circular Gaussian components to fit the visibility data using the task MODELFIT in Difmap. 
The best fit models were determined through an iterative process. We added new model components to an image until the $\chi^2$ value and parameter values of C3 no longer significantly changed. We considered the components consistently observed at similar locations across multiple epochs as the same component. 

In Figure~\ref{fig:series}, we present selected images of 3C~84. While we performed the model-fitting to the entire jet, only the components related to the C3 region are displayed. The Gaussian components highlighted in blue represent C3. Typically, C3 is fitted well with a single Gaussian component as it presents the brightest region in the downstream part of the jet. In cases that C3 exhibits complex morphology, however, additional Gaussian components are required to better represent it. In such cases, multiple Gaussian components are assigned to C3, and we obtained the weighted average position of these components using their flux densities. This average position is used as the position of C3. This method has been used to track the bulk speeds of the M87 jet \citep{park19b}. In addition to C3, multiple subcomponents are detected, originating in the C3 region. We label them with K. 

C3 is identified throughout all epochs and exhibits a continuous southward motion. It is the farthest component from the jet base until the first new jet subcomponent, referred to as K1, emerges from the eastern side of C3 in mid-2017. K1 is cross-identified in 25 consecutive epochs until 2020 February, after which it dissipates. This appears to be the same component identified by previous studies at similar locations. \cite{kino21} identified the component, which they denoted as FW, emerging at the eastern side of the C3 region in 2017 June. They tracked its movement since it was $\sim$3 mas away from the core in 2017 June, until it disappeared at a distance of $\sim$4.5 mas from the core after 2019 March. This is consistent with our investigation. Likewise, \cite{weaver22} identified the component at a similar location and denoted it as C1a, but they claimed that they identified it from 2013.15 until their last epoch obtained in 2018.94. However, we do not find strong evidence in our analysis that this component is connected to a component that existed before its first identification in 2017. To be conservative, we consider K1 started in 2017.

Interestingly, we have identified three more subcomponents that take the appearance of being launched from C3. The second new subcomponent, K2, emerges from the western side of C3 in 2019 October and moves in a southeast direction. The subsequent new subcomponent, K3, also emerges from the western side of C3 in 2021 March and follows a similar path to K2. The last subcomponent, K4, emerges from the eastern side of C3 in 2021 November. We calculated the apparent speed $\beta_{\rm app}$ for these components detected in multiple consecutive epochs.

\subsection{WISE}
The WISE package decomposes the jet structure into a set of two-dimensional significant structural patterns (SSP) using the segmented wavelet decomposition (SWD) method. The algorithm works at multiple size scales, and performs a multiscale cross-correlation (MCC) to cross-identify SSPs detected in multi-epoch images. In the MCC algorithm, the displacement of a component obtained at a larger scale is used to constrain the displacements of its child SSPs at smaller scales. Such series of processes allows for automatic and objective cross-identification of jet components and provides jet velocity fields at multiple scales. For this reason, it has been used to investigate the detailed kinematics of jets with complex structures such as M87 \citep{mertens16, park19b} and 3C~264 \citep{boccardi19} as well as 3C~84 \citep{hodgson21}. However, it is important to consider that the results from WISE could vary depending on the initial parameters such as a size scale, correlation threshold, or significance threshold. Therefore, it is crucial to set parameters suitable for the characteristics of the jet features to be investigated. 

The minimum scale should be set to a sufficiently small value to resolve fine structures such as the new subcomponents from K1 to K4 identified with the MODELFIT analysis. At the same time, the minimum scale should be larger than the smallest decomposable size with the VLBA at 43 GHz. In summary, we initially convolved all images with the average beam size of 0.32$\times$0.15 mas with P.A. of 2.52 degrees. We then decomposed SSPs using the SWD on spatial scales of 0.08 and 0.16 mas and amended them with the intermediate wavelet decomposition (IWD) on scales of 0.12 and 0.24 mas. 
Such investigation at multiple scales is necessary because we cannot know in advance which size scale will best represent the components we observed. 

We set a high significance threshold of 12$\sigma$ to leave only the most evident jet subcomponents. We applied the MCC algorithm with a tolerance factor of 1.4 and a correlation threshold of 0.65. We calculated the velocity of an SSP only when it is cross-identified in at least five consecutive epochs. The velocities of the jet components were calculated toward the average direction of each component. The velocity search windows are set to [$-$1, 1] mas yr$^{-1}$ in $x$ and [$-$3.0, 1.0] mas yr$^{-1}$ in $y$. 

Tracking the bulk motion of C3 using WISE is challenging due to its complex morphology. At size scales smaller than 0.36 mas, the SSPs corresponding to C3 is divided into multiple SSPs. Similar behavior was reported by \citet{hodgson21}. On the other hand, at scales larger than that, C3 and the new subcomponents are blended into a single SSP. Therefore, we used MODELFIT for tracking the motion of C3, while using both MODELFIT and WISE to track the motions of the subcomponents. 

\begin{figure}[t!]
\includegraphics[scale=0.6]{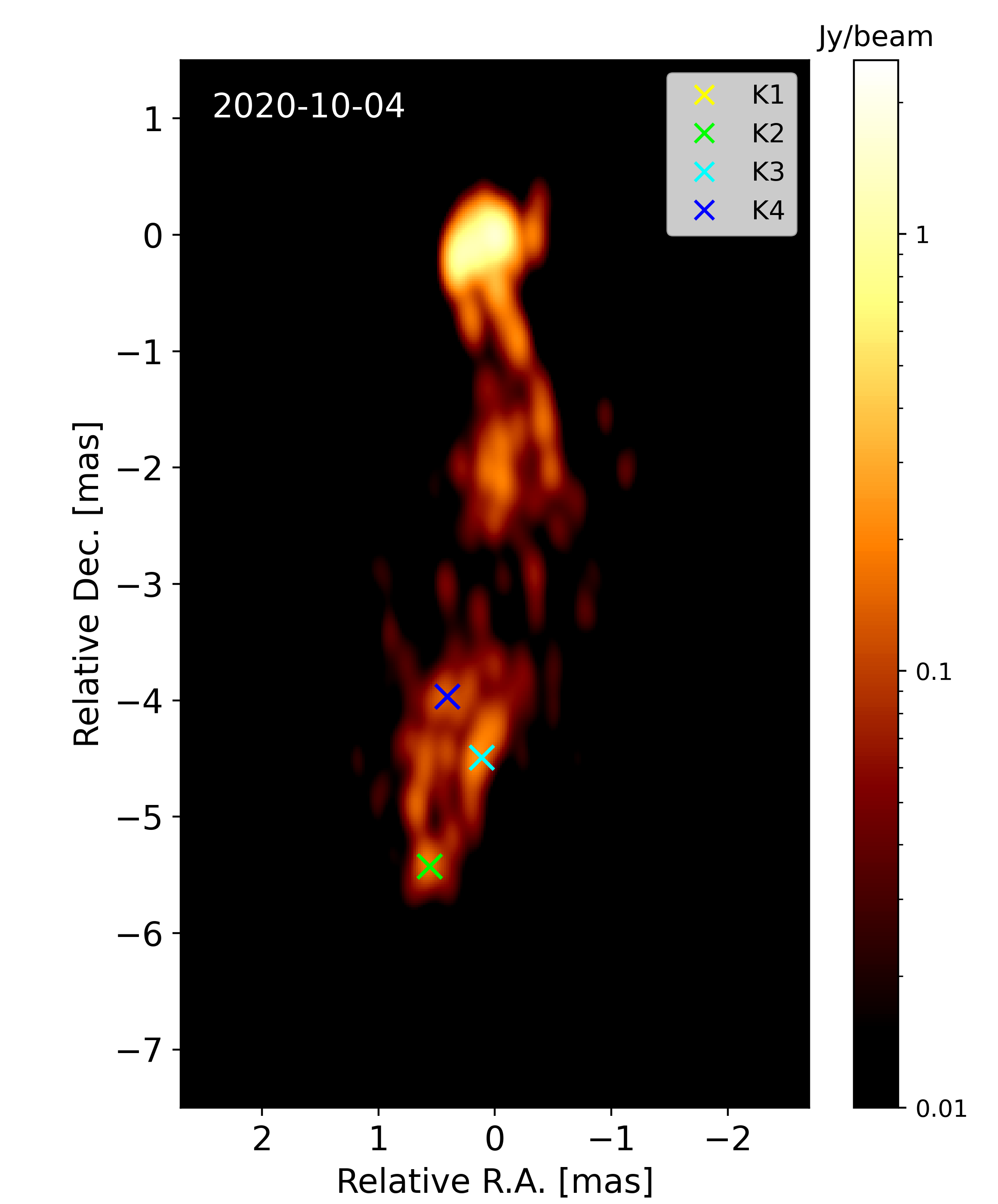}
\caption{A snapshot of an animation illustrating the evolution of the entire jet and the SSPs corresponding to the new subcomponents. The complete animation covering the period from 2010 to 2022 is available online as supplementary material and on YouTube at the following link: \url{https://www.youtube.com/watch?v=6oNgi8oj-Yc}. To enhance the distinction of K1-K4 from the background, we use colors for each subcomponent different from those used in the other figures in this paper. 
\label{fig:movie}}
\end{figure}

\subsection{Cross-identification of new subcomponents}

We found that the displacements of the SSPs at the scale of 0.16 and 0.24 mas are in good agreement with the changes in the contours of the CLEAN maps and the displacements of the MODELFIT components. In Figure~\ref{fig:overplot_enlarged}, we show the SSPs at a scale of 0.16 mas and the MODELFIT components over the CLEAN images of the end of the jet. The epochs, apparent velocities, and direction obtained with each method are given in Table~\ref{tab:summary}. In addition, we have also created an animation displaying the evolution of the jet and the SSPs corresponding to the new subcomponents, K1-K4. While the full movie is available online as a supplementary material and on YouTube, we provide a representative snapshot in Figure~\ref{fig:movie}. The SSPs at a scale of 0.24 mas are shown in Appendix~\ref{appendix_a}. 

The first epoch at which K1 is identified with WISE at a scale of 0.16 mas is 2017.43, whereas it is first identified with MODELFIT in 2017.85. Subsequently, the positions of K1 obtained with both methods are consistent until 2019.80.

\begin{figure*}[ht!]
\centering 
\includegraphics[scale=0.5]{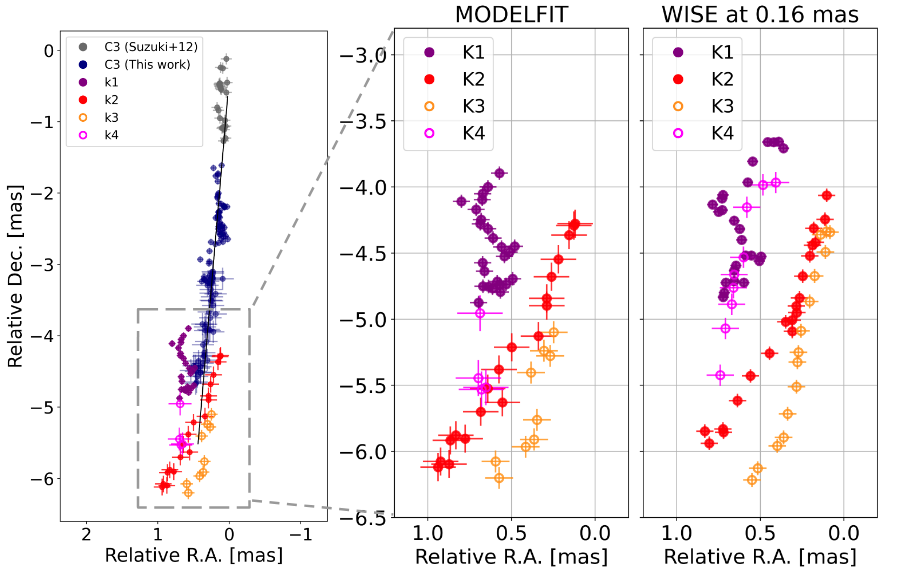}
\caption{Left: All the trajectories obtained with MODELFIT are shown. C3, K1, K2, K3, and K4 are colored navy, purple, red, orange, and magenta. The locations of C3 colored gray are taken from \citet{suzuki12}. The black solid line represents the linear fit to the positions of C3. Right: Enlarged trajectories of the new subcomponents obtained with MODELFIT and WISE are shown. 
\label{fig:trajectory}}
\end{figure*}

K2 is first identified with WISE in 2019.50 when it is located within the region of C3. The emergence of K2 becomes visually apparent in 2019.76, as a distinct localized feature developing in the southwest of C3 and starts to move towards the south. This is the first epoch in which we could confidently put a Gaussian component on K2. In the next epochs, both MODELFIT and WISE consistently track its movement in the southeast direction. K2 is not clearly identified by eye in the time range of 2020.34 to 2020.42 due to its blending with C3, while WISE successfully tracks the movement of K2 during this period. This strengthens our confidence that the component appearing in the southwest of C3 since 2020.51 are indeed the same as K2 identified at similar locations before 2020.34. WISE stops tracking it in 2021.7, while the MODELFIT keeps tracking it until 2022.14, as we still see a localized feature at a similar location. 
K3 is first identified by WISE in 2020.42 within the MODELFIT component for C3. After wobbling within C3 for several months, it begins moving toward the south in late 2020 and is identified by MODELFIT as well since 2021.21. 
WISE begins tracking the last new component K4 in 2020.76 when it is located within C3, while we put the first Gaussian component to K4 in 2021.85 when it begins to move out of C3. WISE stops tracking it after 2022.1, while we put Gaussian components to one more epoch as we still see a localized feature at a similar location. Trajectories and displacements of all the components are shown in Figure~\ref{fig:trajectory} and Figure~\ref{fig:mjd_dist}.

As such, the emergence of four new subcomponents are confirmed with both WISE and MODELFIT analysis. This demonstrates the robustness of the identification of the new subcomponents. \\ 

\subsection{Apparent velocities of C3 and K1-K4}\label{subsec:vapp}

3C~84 started to be monitored with the VLBA-BU-BLAZAR program in 2010 November. To cover the earlier time range of C3, we utilized the positions of C3 reported by \citet{suzuki12}, which were obtained using the VLBA at 43 GHz between 2003 November and 2008 November. We calculated their position uncertainties using the same method applied to the BU data, as described below, in order to maintain consistency between the two datasets. 

For the data obtained in 2010-2018, we identified the core as the brightest component in the upstream flow of the jet, which we label C1. This is consistent with other studies that also labeled this region as C1 \citep[e.g., ][]{nagai14}. In 2019, the structure of the C1 region becomes complex due to the ejection of a bright component, as reported by \citet{park23}. This results in the brightest component being either the new component or the actual core, depending on the epoch. Following \citet{park23}, we identified the western component as C1 in the data obtained since 2019. After defining C1 as described above, we measured the position of each component relative to C1.


There is no straightforward way to estimate the position uncertainty of each component. We followed a method described in \citet{homan01}. We initially set the position uncertainty to be equal for all the data points of each component, and performed a linear regression to each component, assuming that their apparent velocities are constant in time. After that, we uniformly re-scaled the position uncertainties of each component until the reduced $\chi^2$ reaches $\sim$1 (see Appendix~\ref{appendix_b} for details). 

\begin{figure*}[ht!]
\centering 
\includegraphics[width=\textwidth]{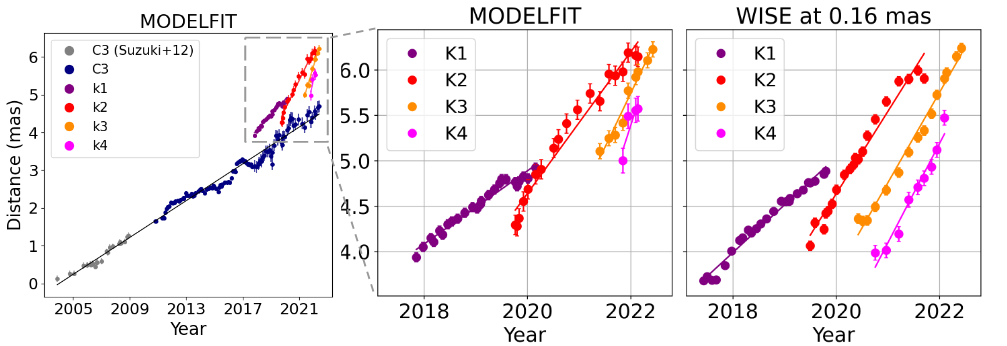}
\caption{Distance from the core as a function of time for each jet component. C3, K1, K2, K3, and K4 are colored navy, purple, red, orange, and magenta, respectively. Gray dots represent the locations of C3 taken from \citet{suzuki12}. The solid lines represent the linear displacements of each component obtained from the best-fit analysis. On the left, we present the displacement of each jet component obtained with MODELFIT for the entire time range of C3 (from 2003 November to 2022 June). In the middle, the displacements of the new subcomponents obtained with MODELFIT are displayed. The displacements of the new subcomponents obtained with WISE are displayed on the right.} 
\label{fig:mjd_dist}
\end{figure*}

The apparent velocity of C3 seems to exhibit abrupt changes occasionally (see Figure~\ref{fig:mjd_dist}). If we assume that the apparent velocity is constant over the entire time range, it would lead to a larger position uncertainty. To avoid this, we selected a time range between 2012 January and 2016 March where the apparent velocity of C3 remains relatively constant over $\sim$4 years. We applied this position uncertainty to the other periods, except the period between 2017 November and 2022 June. During this period, the position scatter of C3 is larger due to the emergence of new subcomponents, the growing complexity of its morphology, and the decay of its brightness. For these reasons, we separately calculated the position uncertainty of C3 for the period between 2017 November and 2022 June.  

Table~\ref{tab:summary} summarizes the direction of propagation and apparent velocity in that direction for each component. The difference in the apparent velocities of the new subcomponents measured by the two different methods is larger than the measurement uncertainties. This difference arises from the fact that the number of epochs for which each subcomponent is identified differs depending on the methods used (see Appendix~\ref{appendix_c} for details). Interestingly, K3 and K4 exhibit superluminal speeds regardless of the method used to measure their apparent velocities. In addition, there appears to be a trend towards faster apparent speeds in the more recent data. 

\begin{table*}[ht]
    \centering 
    \caption{Jet Features \label{tab:summary}}
    \begin{tabular}{cccccc}
        \hline
        Method & Name & Epoch & $\beta_{\rm app}$ & $\mu_{\rm app}$ (mas yr$^{-1}$) & P.A. ($^\circ$) \\
        \hline 
        \multirow{5}{*}{MODELFIT} & {C3} & 2003.89 - 2022.64 & $0.259\pm0.003$ & 0.233 $\pm$ 0.002 & $176.5\pm0.4$ \\ \cline{2-6}
        & K1 &  2017.85 - 2020.16 & $0.46\pm0.02$ & $0.41\pm0.02$ & $185.5\pm2.2$  \\
        & K2 &  2019.76 - 2022.14 & $0.91\pm0.06$ & $0.82\pm0.05$ & $155.9\pm2.5$ \\
        & K3 &  2021.21 - 2022.64 & $1.35\pm0.14$ & $1.22\pm0.13$ & $164.4\pm4.3$ \\
        & K4 &  2021.85 - 2022.14 & $1.99\pm1.09$ & $1.79\pm0.98$ & $182.9\pm21.9$ \\
        \hline 
        \multirow{4}{*}{WISE (0.16)}  & K1 &  2017.43 - 2019.80 & $0.59\pm0.03$ & $0.53\pm0.03$ & $171.2\pm2.0$ \\ 
        & K2 &  2019.50 - 2021.70 & $1.05\pm0.06$ & $0.95\pm0.05$ & $158.3\pm2.2$ \\
        & K3 &  2020.42 - 2022.42 & $1.11\pm0.06$ & $1.00\pm0.06$ & $168.6\pm2.1$ \\
        & K4 &  2020.76 - 2022.10 & $1.22\pm0.14$ & $1.10\pm0.13$ & $167.7\pm4.7$ \\
        \hline
    \end{tabular}
\tablecomments{Jet features of each component identified with MODELFIT and WISE at a scale of 0.16 mas. Due to the complex morphology of C3, we only used MODELFIT to track its motion (see Section~\ref{sec:analysis} for more details.)}
\end{table*}

\section{Discussion} \label{sec:Discussions}
\subsection{Kinematics of C3}  \label{sec:c3}
Our findings reveal that the average apparent velocity of C3 over the past 20 years is $\beta_{\rm app} = 0.259\pm0.003$. This is within the range of the previously reported values of $\beta_{\rm app}=0.23\pm0.01$ from 2003 November to 2008 November \citep{suzuki12}, $\beta_{\rm app}=0.23\pm0.06$ from 2007 October to 2009 April \citep{nagai10}, $\beta_{\rm app}=0.27\pm0.02$ from 2007 October to 2013 December \citep{hiura18}, $\beta_{\rm app}=0.25\pm0.01$ from 2010 November to 2018 December \citep{weaver22}, and $\beta_{\rm app}=0.2-0.3$ from 2013 January to 2016 December \citep{hodgson18} (Table~\ref{tab:vapp}). This suggests that the apparent speed of C3 remained nearly constant for the past 20 years. In addition, the global direction of C3 also remained nearly constant (Figure~\ref{fig:trajectory}). 


It is interesting to note that previous studies have reported either apparent acceleration \citep{suzuki12} or deceleration \citep{kino18, kino21} of C3 at various times. Indeed, C3 does exhibit abrupt changes in its apparent speed several times but its apparent velocity remains constant when averaged over longer times scales (see Figure~\ref{fig:mjd_dist}). Additionally, the global direction of C3 remains nearly constant over the past 20 years (see Figure~\ref{fig:trajectory}), while exhibiting motion in alternating directions transverse to the global direction on short timescales \citep{suzuki12, hiura18, kino18, kino21}. This can be interpreted as the result of precession of the jet. \citet{britzen19} and \citet{dominik21} suggest that the direction of the jet changes over time with a precession period of 30$\sim$40 years. This is also seen in our images (see Figure~\ref{fig:series}). The direction of the jet near the core is toward the southeast in early 2010s, while it is changed toward the southwest in 2020s. In addition, limb-brightening is enhanced in the direction of precession, which is commonly observed in three-dimensional relativistic hydrodynamical simulations of precessing jets \citep[e.g.,][]{nawaz16}. This suggests that precession would introduce transverse motion to C3, leading to the variation in the viewing angle of C3 on short timescales. This could in turn lead to the variation of the apparent velocity on short timescales. Furthermore, precession provides a possible explanation on the ejection of the new subcomponents. This will be discussed in Section~\ref{sec:precession}. 
Another explanation for the small, rapid transverse motions of C3 is that the jet stream is propagating through inhomogeneous media, in which the jet termination region encounters a variation in density, and, thus, ram-pressure. Further consequences of jet-cloud interactions that may be consistent with the ejection of new subcomponents are discussed in Section~\ref{sec:jetcloud}.

\begin{table}[t]
    \centering
    \setlength{\tabcolsep}{3pt}
    \caption{Summary of the apparent velocities of C3}
    \begin{tabular}{c c c c}
    \toprule
        Period & $\beta_{\rm app}$ & $\mu_{\rm app}$  (mas yr$^{-1}$) & Ref. \\
        \midrule 
        2003 Nov. - 2008 Nov. & 0.23 $\pm$ 0.06 & 0.21 $\pm$ 0.05 & (1) \\
        2007 Oct. - 2009 Apr. & 0.23 $\pm$ 0.01 & 0.20 $\pm$ 0.01 & (2)  \\
        2007 Oct. - 2013 Dec. & 0.27 $\pm$ 0.02 & 0.23 $\pm$ 0.02 & (3) \\
        2010 Nov. - 2018 Dec. & 0.25 $\pm$ 0.01 & 0.23 $\pm$ 0.01 & (4) \\
        2013 Jan. - 2016 Dec. & 0.2 - 0.3 & 0.17 - 0.26 & (5) \\
        2003 Nov. - 2022 Jun. & 0.259 $\pm$ 0.003 & 0.233 $\pm$ 0.002 & (6) \\ 
    \bottomrule
    \end{tabular}
    \tablecomments{(1) \citet{suzuki12}, (2) \citet{nagai10}, (3) \citet{hiura18} (4) \citet{weaver22} (5)\citet{hodgson18} (6) this work.
    }
    \label{tab:vapp}
\end{table}

\subsection{Intrinsic velocity of the new subcomponents}
We report the detection of superluminal motions within 3C~84. 
In particular, the subcomponent K4 exhibits an apparent speed of 1.99$\pm$1.08c as detected by MODELFIT, albeit with a large uncertainty. WISE also detected it with a lower but more precise apparent speed of 1.22$\pm$0.14c. The faster motion with a larger uncertainty obtained with the MODELFIT analysis could be due to insufficient time range for sampling. We consider the lower speed obtained with WISE to be more reliable.

We examined whether changes in the viewing angle could explain the larger apparent velocities observed in the new subcomponents compared to C3. Assuming that the intrinsic velocity of both C3 and new subcomponents are equal, we investigated possible viewing angle ranges for each subcomponents that could account for their higher apparent velocities. The apparent velocity can be expressed as follows:\
\begin{equation}\label{eq:vapp}
    \begin{aligned}
    \beta_{\rm app} & = \frac{\beta_{\rm int} \sin \theta_{\rm view}}{1-\beta_{\rm int} \cos \theta_{\rm view}} 
    \end{aligned}
\end{equation}

\noindent where $\beta_{\rm int}$ is the intrinsic speed of the components in units of c and $\theta_{\rm view}$ is the viewing angle that represents the angle between the direction of the jet and our line of sight. 

The reported viewing angles of 3C~84 are largely divided into two groups. One group uses the spectral energy distribution (SED) analysis using a simple one-zone synchrotron self-Compton (SSC) model \citep{abdo09, aleksic14, tanada18} or a `spine-sheath' model \citep{tavecchio14}. The viewing angles derived from these methods fall within the range of $6^\circ-30^\circ$. The other group uses the ratio between the lengths of the approaching and receding jets. The range of the viewing angle of the inner jet obtained from this method is $39^\circ-65^\circ$ depending on the location of the black hole \citep{fujita17}.
This is consistent with the range of the viewing angles for the outer jet located 13$-$14 mas south downstream of the inner jet, which is $30^\circ-55^\circ$ \citep{walker94, asada06}. Recently, \citet{oh22} obtained an upper limit of the viewing angle in C1 of $\lesssim35^\circ$ using the ratio between the jet/counter-jet flux densities. This upper limit for C1 is consistent with the values obtained from the SED analysis. They suggested that the discrepancy in the viewing angle obtained by different methods could be due to the difference in the viewing angles between C1 and C3. Considering the discussions above, we assume that the average viewing angle of C3 would be in the range of 30$^\circ-65^\circ$. 

\begin{figure}[t!]
\centering 
\includegraphics[scale=0.6]{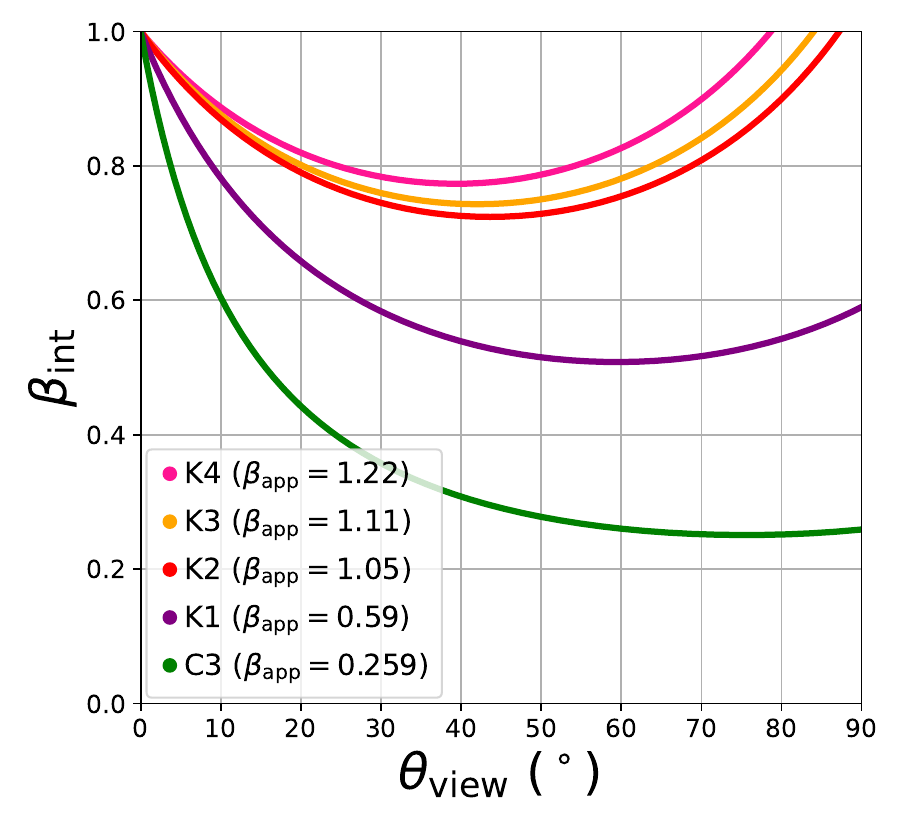}
\caption{Solid lines in green, purple, red, orange, and magenta represent possible pairs of $\beta_{\rm int}$ and $\theta_{\rm view}$ that account for $\beta_{\rm app}$ of C3, K1, K2, K3, and K4, respectively. 
\label{fig:pspace}}
\end{figure}

Figure~\ref{fig:pspace} shows possible pairs of intrinsic velocity and viewing angle that corresponds to the observed apparent velocities of C3 and the new subcomponents. If the difference in apparent velocities between C3 and K4 solely stems from the difference in their viewing angles, the intrinsic velocity of C3 must have a lower limit of $\beta_{\rm int}=0.77$ and the viewing angle of C3 must have an upper limit of 4.4$^\circ$. However, this is in tension with the viewing angle range of $30^\circ-65^\circ$ as derived from the jet/counter-jet analysis \citep{walker94, walker00, asada06, fujita17}. Therefore, faster apparent speeds of the new subcomponents cannot be explained by the change in the viewing angle alone. This indicates that the new subcomponents are intrinsically faster than C3. \\

\subsection{Formation mechanisms for the new subcomponents}

Here we explore possible mechanisms that might produce new subcomponents faster than C3. We first consider two mechanisms of hydrodynamic origin: precession, and jet-cloud interactions. Both produce split jet beams and multiple hotspots, and do not require an acceleration mechanism for the new subcomponents to move faster than C3 because the secondary jet streams tend to propagate through more diffuse media. These two mechanisms are, however, very different in the way the multiple subcomponents and hot spots are produced: precession involves jet-internal hydrodynamic instabilities and interactions with a turbulent backflow in the cocoon, while jet-cloud interactions depend solely on the direct scattering of jet streams in a clumpy ambient medium. As a third possibility, we also consider magnetic reconnection as a mechanism that accelerates fast subcomponents.

\subsubsection{Hydrodynamic instabilities due to changes in jet direction} \label{sec:precession}

Recent hydrodynamic simulations of precessing jets by \citet{horton23} suggest that precession can lead to the formation of multiple hotspots in Fanaroff \& Riley class II (FR II) radio jets \citep{fr74}. While the following discussion focuses on precessing jets, we note that any process leading to a change in jet direction would yield similar results. 
A stationary jet interacting with the smooth interstellar medium would, in general, not produce subcomponents, since the jet remains stable up to the head of the lobe, where it is shocked and produces a backflow. If the jet direction changes, however, it forms a more extended lobe, and the main jet stream tends to be deflected at an oblique angle close to sides of the lobe, enhancing instabilities in the main jet stream. The deflection itself can lead to multiple oblique shocks and hotspots, and the instabilities can perturb the jet significantly such that it splits into two or more sub-streams, or such that it creates a cluster of shocked plasmoids. The splitting of the main jet stream may be enhanced due to the interaction of the jet stream with the backflowing plasma as its direction continuously sweeps around the turbulent lobe. The deflected or split jet streams or plasmoids may tend to propagate at a faster speeds into the diffuse cocoon, creating fast-moving hotspots, compared to the speed of the original hotspot (C3), which is limited by momentum balance with the ambient interstellar medium. The secondary hotspots may not be long lived and fade away fairly quickly if the secondary jet streams and plasmoids become disconnected from the main jet stream. Similar phenomena have also been observed in other simulations of precessing jets \citep{nawaz16}. 

The mechanisms described above of the formation of multiple hotspots could correspond to the ejection of new subcomponents from C3. Previous studies suggested that the jet of 3C~84 either precesses with a period of 30$\sim$40 years \citep{britzen19, dominik21} or reoriented \citep{park23}. In either case, any rapid change in jet direction would lead to the creation of an extended lobe region, causing C3 to interact with the lobe boundary at an oblique angle and cause the jet to repeatedly interact with turbulent back-flowing plasma. Indeed, recent RadioAstron space-VLBI observations at 5~GHz have detected a cocoon-like structure around the jet \citep{savolainen23}. In addition, back-flowing plasma has also been identified on the western side of C3 \citep{nagai14, nagai16}, which is also visible in our images from 2010 to 2017 (see Figure~\ref{fig:series}). Notably, the ejection of the first subcomponent K1 is accompanied by the disruption of the initial hotspot \citep{kino21}. This suggests that K1 would be a secondary hotspot that split from the initial hotspot through the interaction between the jet and the lobe boundary and the turbulent backflow. Additionally, $\beta_{\rm app}=0.46\pm0.02$ of K1 obtained with MODELFIT is almost identical to $\beta_{\rm app}=0.44\pm0.08$ of C3 obtained with the same method during the same period, while the subsequent new subcomponents exhibit much faster apparent speeds. The fast speeds may be due to K2-K4 propagating through the diffuse turbulent jet cocoon within the lobe expanded by K1, while C3 and K1 are pushing against the denser ambient interstellar medium. 

It is noteworthy that the subcomponents emerge only after 2017. This would be related to the rapid change of the jet direction in 2015. The jet direction near the core was toward the southwest from its first identification from 2003 \citep{suzuki12} until 2014 (Figure~\ref{fig:series}), spanning nearly 10 years. In 2015, the jet direction near the core began to rapidly shift from southwest to southeast. The time it takes for this rapid directional change to propagate to C3 is consistent with the formation of K1-K4 only after 2017. 

The hydrodynamical simulations of precessing jets \citep[e.g.,][]{nawaz16, horton23} have been conducted based on the physical quantities of AGN jets on tens of kiloparsec-scales, while 3C~84 is a jet on parsec-scales. The production of multiple hotspots through rapid changes in jet direction depends sensitively on jet properties such as jet power, bulk Lorentz factor, internal Mach number, and the speed of jet direction change. To confirm the feasibility of this scenario, hydrodynamic simulations tailored to 3C~84 are required.

\subsubsection{Jet-cloud collision}\label{sec:jetcloud}
Recent studies suggest the possibility that the jet of 3C~84 collided with dense gas clouds \citep[e.g.,][]{kino18, kino21, park23}. Numerical simulations of collisions between non-precessing jets and clouds may provide hints to understand the observational results in this work. When non-precessing jets collide with clouds, a portion of the jet flow is deflected or split with a change in direction \citep[e.g.,][]{deG_DalPino99, wagner12}. This process will create a slow-moving hotspot around the region where the jet is hitting the cloud, and faster-moving hotspots will be formed farther out near the tip of the deflected jet. If a jet-cloud collision occurred in 2017-2018 \citep{kino21}, then the deflected jet could potentially explain the ejection of the new subcomponents. In this scenario, the subcomponents would be faster than C3 as they are propagating through more diffuse interstellar media, while C3 is formed at the point where the jet impacts the cloud. This difference in velocity would be particularly large if the deflected jet streams flow out of a denser region of clouds into the open interstellar medium. 
The fact that K1-K4 are relatively aligned with the direction of C3 may constrain the size of clouds that are deflecting or scattering the primary jet stream. If the clouds are very large compared to the width of the jet stream, the deflection angles with respect to the main jet stream may be too large \citep[e.g., Figure 9 in][]{wagner12}. The strong alignment of the direction of K1-K4 would point to interactions of the jet stream with clumps whose size are of order the jet diameter \citep[e.g., Figure 1, middle row, in][]{wagner12}.

One diagnostic indicator for distinguishing between the jet-cloud collision and jet precession scenarios could be the presence of shock tracers in dense gas in the former. The interaction of a jet with dense clouds in the ISM would drive radiative shocks into these clouds, which could be seen in emission lines such as [O\,\textsc{iii}] from radiatively cooling post-shock layers. On the other hand, the precession model does not feature interactions with the dense clouds in the ambient medium and would preclude such emission lines. While the optical spectrum of 3C84 does exhibit [O\,\textsc{iii}] and other narrow emission lines, the existence of broad emission lines suggests these narrow lines also originate from the gas ionized by the accretion disk \citep[e.g.,][]{son12, punsly18}. Spectroscopy with milliarcsecond-scale spatial resolution will be necessary to ascertain whether any [O\,\textsc{iii}] emission lines originate at C3. 

\subsubsection{Magnetic reconnection}
Magnetic reconnection occurs when magnetic field lines undergo a topological rearrangement, resulting in a rapid release of magnetic energy. This process takes place when magnetic field lines of opposite polarity converge towards the reconnection plane, causing them to annihilate. As a result, magnetic energy is released that heats the plasma and accelerates particles, which leads to $\gamma$-ray emission. The magnetic pressure gradient and magnetic tension perpendicular to the magnetic field lines contribute to the bulk acceleration, leading to the formation of mini-jets \citep[see][for more details]{giannios09, giannios13}. This mechanism has been suggested to explain the fast TeV flaring events observed in several blazars \citep[e.g.,][]{aharonian07, aleksic11, jorstad13}. 

Rapid TeV flares have also been detected in 3C~84 \citep[e.g.,][]{aleksic12, aleksic14, magic18}, and C3 has been considered as one of possible origins of the gamma-ray flares \citep{nagai16, hodgson18, hodgson21}. In particular, \citet{hodgson21} suggested that magnetic reconnection at C3 could be the physical mechanism responsible for some $\gamma$-ray flares in 3C~84, leading to the formation of mini-jets \citep{giannios09, giannios13}. In this scenario, the physical interpretation for the ejection of K1 to K4 and their related increase of intrinsic jet speeds is that they are the mini-jets. 
However, it is challenging to pinpoint the ejection moment of the new subcomponents at C3, as the first identification epoch of each component varies depending on the methods or decomposition scales. This makes it difficult to specify which $\gamma$-ray flare corresponds to the ejection of a specific subcomponent.\\ 
Instead, we briefly examine if the observed quantities are consistent with this scenario. Consider C3 that moves radially with bulk Lorentz factor $\Gamma_{\rm j}$, containing new subcomponents with Lorentz factor $\Gamma_{\rm co}$ at angle $\theta'$ with respect to the direction of C3 in the rest frame. The primed/tilded quantities represent the quantities measured in the rest frame. If we assume that the fastest new component K4 leaves the reconnection region C3 with bulk Lorentz factor close to the Alfv\`en speed of the upstream plasma \citep{lyubarsky05}, the observed Lorentz factor of K4, $\Gamma_{\rm em}$, can be expressed as \citep{giannios09} 

\begin{equation}\label{eq:giannios1}
    \Gamma_{\rm em} \sim \Gamma_{\rm j}\sqrt{\sigma} (1+\beta_{\rm j}\beta_{\rm co} \cos{\theta'}) 
\end{equation}




\noindent where $\sigma$ is Poynting-to-kinetic flux ratio. Considering that the directions of K4 is nearly aligned with the global direction of C3 (Table~\ref{tab:summary}), the equation~\ref{eq:giannios1} can be approximated as 

\begin{equation}\label{eq:reconnection}
    \Gamma_{\rm em} \sim \Gamma_{\rm j}\sqrt{\sigma}\times1.37
\end{equation} 

\noindent by assuming $\cos{\rm \theta'}\sim0$ and using $\beta_{\rm j}\sim0.259$ and $\beta_{\rm co}\sim1.41$. If we assume the minimum Lorentz factor for K4 of $\Gamma_{\rm em}\sim1.7$, and minimum Lorentz factor for C3 of $\Gamma_{\rm j}\sim1.04$ assuming $\beta_{\rm int}=0.25$, this gives $\sigma\sim1.5$. This point towards a magnetized jet which would be expected in a reconnection scenario \citep[see Figure 6 in][]{kataoka05}. If the energy release could consistently occur in the same direction as C3, magnetic reconnection could explain the formation of the subcomponents. 

\subsection{Components launched from C1?}
Lastly, we consider the possibility that K1-K4 are related to specific components identified near C1. By assuming that their apparent speeds have remained nearly constant, we extrapolated the trajectories of K1-K4 obtained with WISE at a scale of 0.16 mas to C1 and found that the ejection times correspond to 2010 May, 2014 November, 2016 March, and 2017 April for K1, K2, K3, and K4, respectively. During this period, however, only two components were identified as being ejected from C1 \citep{weaver22, paraschos22}. This suggests that K1-K4 could be matched to the components identified near C1 only if some of K1-K4 were ejected before 2010 with slower initial speeds but accelerated between C1 and C3. Recent observations of nearby radio galaxies such as M87 and NGC~315 suggest that the jets are gradually accelerated from subluminal to superluinal speeds as they approach the Bondi radius \citep[e.g.,][]{asada14, mertens16, walker18, park19b, park21}. However, such gradual acceleration has not been found in 3C~84 \citep{hodgson21}. Instead, one component was observed to be abruptly accelerated from 0.32c to 1.60c near C1 \citep{park23}. While the physical mechanism for such abrupt acceleration is unclear, if it could have occurred for K1-K4, then the components launched from C1 could potentially serve as the origin of K1-K4. \\

\section{Conclusions} \label{sec:Conclusions}
We have investigated the kinematics of C3 and the new subcomponents that emerged from C3. We have found that the average apparent velocity of C3 is $\beta_{\rm app}=0.259\pm0.003$ over the past 20 years. Since 2017, four new subcomponents K1-K4 took the appearance of being launched from C3. The apparent speeds of all four subcomponents are larger than that of C3, and K2-K4 exhibit faster apparent speeds than K1.  

We have examined whether the apparent speeds of the subcomponents could exceed that of C3 with the change in the viewing angle alone. This requires the viewing angle of C3 to be smaller than $4.4^\circ$, otherwise the change in viewing angle alone cannot explain the faster apparent speeds of the new subcomponents. A viewing angle smaller than $4.4^\circ$ is, however, inconsistent with almost all estimates of the jet orientation. This suggests that the new subcomponents are intrinsically faster than C3.   

We have discussed possible mechanisms that could explain the ejection of relativistic subcomponents from C3. We have concluded that rapid changes in jet direction, jet-cloud collision, and magnetic reconnection could be responsible for the ejection of the new subcomponents that can propagate faster than C3. It is also possible that the subcomponents originate from C1, in which case the nature of their formation and acceleration is unclear. \\

\begin{acknowledgments}
\noindent We thank an anonymous referee for constructive and useful comments that improved this paper. M. Kam acknowledges A. Marscher and S. Jorstad for useful discussions on the BU data. We acknowledge financial support from the National Research Foundation of Korea (NRF) grant 2022R1F1A1075115. M.Kino is supported by the JSPS KAKENHI Grant number JP22H00157 and JP21H01137. H.N. is supported by the JSPS KAKENHI Grant number JP21H01137 and JP18K03709. 
This study makes use of VLBA data from the VLBA-BU Blazar Monitoring Program (BEAM-ME and VLBA-BU-BLAZAR; http://www.bu.edu/blazars/BEAM-ME.html), funded by NASA through the Fermi Guest Investigator Program. The VLBA is an instrument of the National Radio Astronomy Observatory. The National Radio Astronomy Observatory is a facility of the National Science Foundation operated by Associated Universities, Inc. 
\end{acknowledgments}

%

\vspace{5mm}

\bibliography{ref}{}
\bibliographystyle{aasjournal}

\appendix

\section{WISE at scales of 0.16 and 0.24 mas}\label{appendix_a}

\begin{figure*}[h!]
\centering 
\includegraphics[width=\textwidth]{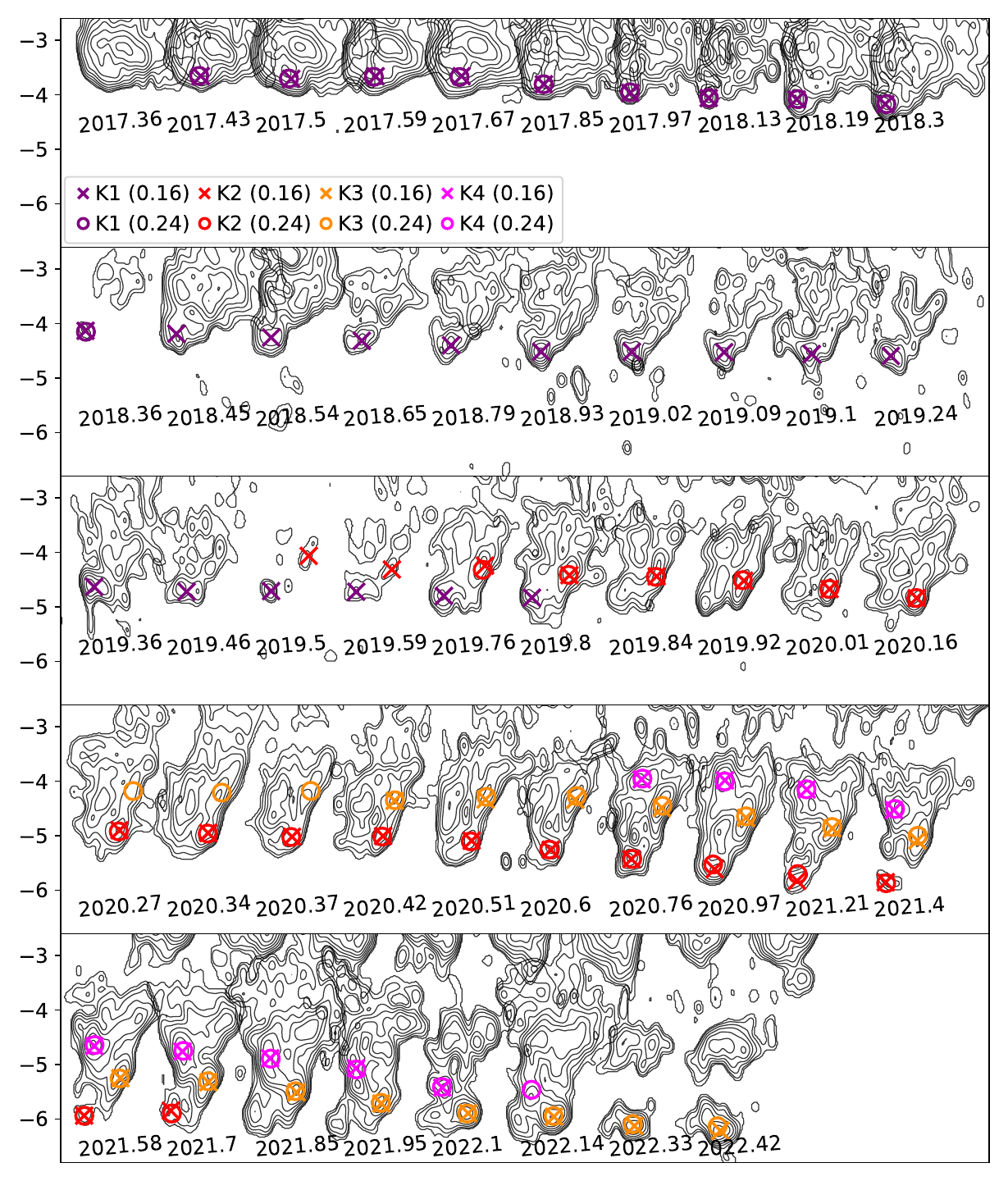}
\caption{The enlarged CLEAN images of the jet termination region of 3C~84 obtained from 2017.36} to 2022.42. SSPs identified with WISE at scales of 0.16 and 0.24 mas are shown in circles and crosses, respectively. K1-K4 are colored purple, red, orange, and magenta, respectively. 
\label{appendix:overplot}
\end{figure*}

\begin{figure*}[ht!]
\centering 
\includegraphics[scale=0.53]{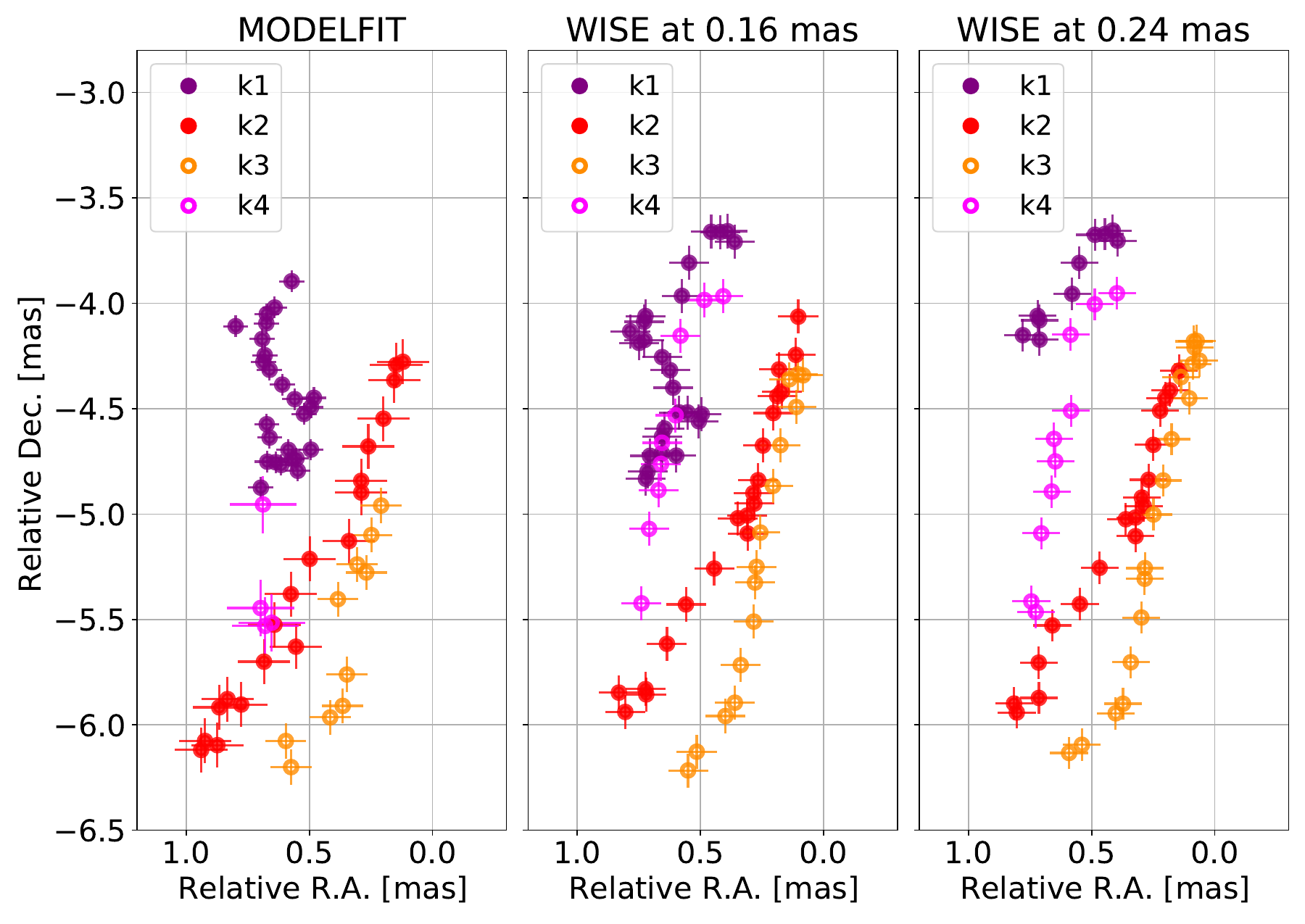}
\caption{On the left, the trajectories of the new subcomponents obtained with MODELFIT are shown. K1-K4 are colored purple, red, orange, and magenta. In the middle and right, the trajectories of the subcomponents obtained with WISE at scales of 0.16 and 0.24 mas are shown, respectively.
\label{appendix:trajectory}}
\end{figure*}

\begin{figure*}[ht!]
\centering 
\includegraphics[width=\textwidth]{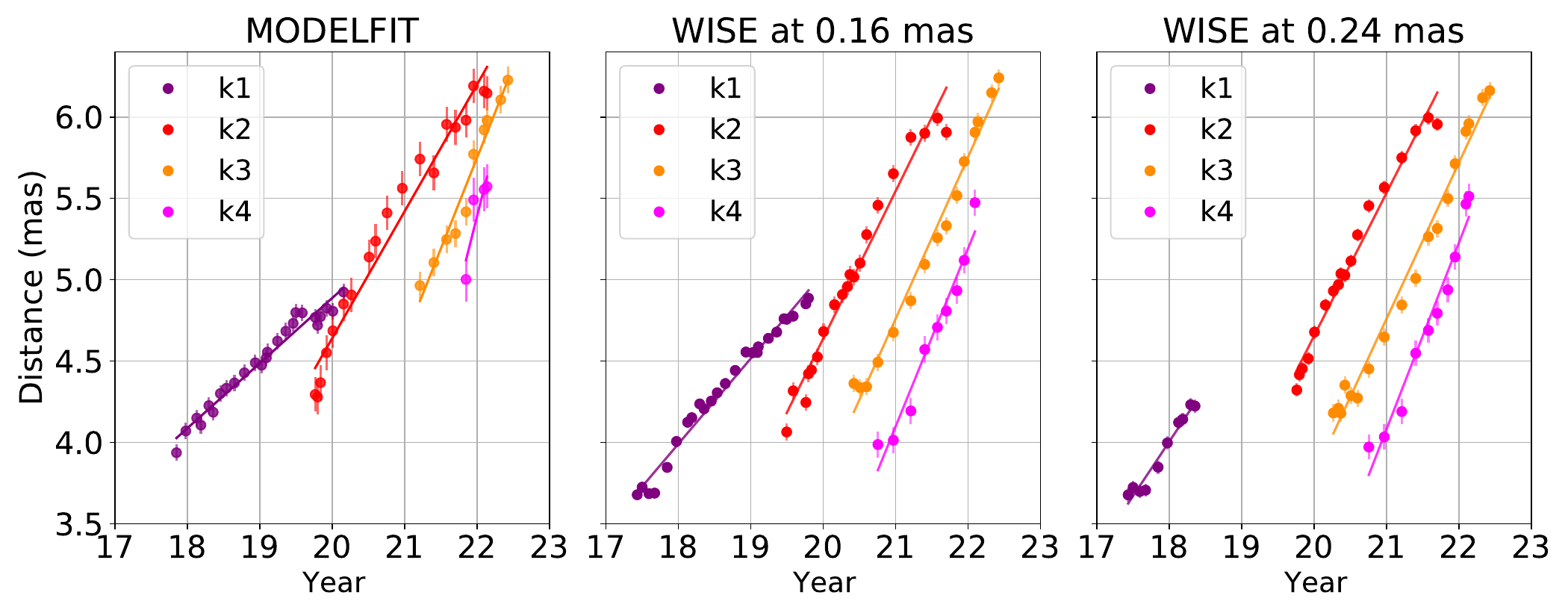}
\caption{Distance from the core as a function of time for each subcomponent. K1, K2, K3, and K4 are colored purple, red, orange, and magenta, respectively. The solid lines represent the linear displacements of each subcomponent obtained from the best-fit analysis. On the left, we present the displacement of each subcomponent obtained with MODELFIT for K1-K4. In the middle and right, the displacements of the subcomponents obtained with WISE at 0.16 and 0.24 mas are displayed, respectively.} 
\label{appendix:mjd_dist}
\end{figure*}

In this appendix, we compare the SSPs corresponding to the new subcomponents K1--K4 obtained at scales of 0.16 and 0.24 mas to further demonstrate the robustness of the detection of the new subcomponents. SSPs at a scale of 0.16 and 0.24 mas are shown in circles and crosses, respectively. In Figure~\ref{appendix:overplot}, the SSPs at the two different scales are plotted over the CLEAN images of the jet termination region of 3C~84. The new subcomponents from K1 to K4 are colored purple, red, orange, and magenta, respectively. The SSPs at both scales consistently identify K1 from 2017.43 to 2018.36. K2 is tracked since 2019.5 and 2019.76 at scales of 0.16 and 0.24 mas, respectively. WISE stops tracking it in 2021.7 at both scales. K3 is tracked since 2020.42 and 2020.27 at scales of 0.16 and 0.24 mas, respectively. From 2020.76 to 2022.1, K4 is consistently tracked at both scales. These results show that all the new subcomponents are cross-identified with WISE at both scales.
Figure~\ref{appendix:trajectory} and Figure~\ref{appendix:mjd_dist} show the trajectories and separation from the core of the new subcomponents obtained from the MODELFIT and WISE at both scales. 

\begin{table*}[b!]
    \centering 
    \begin{threeparttable}
    \caption{Position uncertainty\label{tab:uncertainty}}
    \begin{tabular}{ccccccc}
        \hline
        Method & Name & Epoch & N\tnote{a} & $\theta$ [mas]\tnote{b} & $\sigma$ [mas]\tnote{c} & $\sigma$/beam\tnote{d} \\
        \hline 
        \multirow{6}{*}{MODELFIT} & C3a & 2012.07 - 2016.21 & 40 & 0.65 & 0.039 & 0.12  \\ 
        & C3b & 2017.85 - 2022.42 & 43 & 1.08 & 0.180 & 0.56  \\ 
        \cline{2-7}
        & K1 & 2017.85 - 2020.16 & 25 & 0.34 & 0.050 & 0.16 \\
        & K2 & 2019.76 - 2022.14 & 19 & 0.30 & 0.117 & 0.37 \\
        & K3 & 2021.21 - 2022.64 & 9 & 0.61 & 0.088 & 0.27 \\
        & K4 & 2021.85 - 2022.14 & 4 & 0.31 & 0.161 & 0.50 \\
        \hline 
        \multirow{4}{*}{WISE (0.16)} & K1 & 2017.43 - 2019.80 & 25 & \multirow{4}{*}{0.16} & 0.054 & 0.17 \\ 
        & K2 & 2019.50 - 2021.70 & 20 & & 0.108 & 0.34 \\
        & K3 & 2020.42 - 2022.42 & 15 & & 0.088 & 0.27 \\
        & K4 & 2020.76 - 2022.10 & 9 & & 0.113 & 0.35 \\
        \hline
        \multirow{4}{*}{WISE (0.24)} & K1 & 2017.43 - 2018.36 & 10 & \multirow{4}{*}{0.24} & 0.048 & 0.15 \\ 
        & K2 & 2019.76 - 2021.70 & 18 & & 0.082 & 0.26 \\
        & K3 & 2020.27 - 2022.42 & 18 & & 0.098 & 0.31 \\
        & K4 & 2020.76 - 2022.14 & 10 & & 0.114 & 0.36 \\
        \hline
    \end{tabular}
\tablecomments{Position uncertainties of each component obtained with MODELFIT and WISE at scales of 0.16 and 0.24 mas. \\ 
$^{\rm a}$ Number of the epochs that each component is identified. \\
$^{\rm b}$ The size of each component. \\ 
$^{\rm c}$ Position uncertainty \\
$^{\rm d}$ Position uncertainty relative to the beam size} 
\end{threeparttable}
\end{table*}

\section{Position uncertainty}\label{appendix_b}
The position uncertainties of C3 and K1--K4 are summarized in Table~\ref{tab:uncertainty}. Previous studies, which estimated the position uncertainties using the same method \citep[i.e, the method of][]{homan01}, suggest that the position uncertainties are typically $\sim$20\% of the beam size and tend to be smaller by a factor of $\sim$2 for isolated bright and compact features \citep[e.g.,][]{lister09, lister16, lister19, lister21}. 
Here we investigate the relative magnitude of the position uncertainty for C3 and K1-K4 compared to the beam size. 

The average beam size of all 112 images used for our analysis is 0.32$\times$0.15 mas with a P.A. of 2.52 degrees. Considering that the P.A. of the beam is nearly parallel to the the global jet direction of 176.5$^\circ$, we assumed 0.32 mas as the beam size and obtained the position uncertainties relative to the beam size. As described in Section~\ref{subsec:vapp}, we estimated the position uncertainty of C3 using two different time ranges. For the time range between 2012 January and 2016 March, during which the apparent velocity of C3 remained relatively constant over $\sim$4 years (denoted as C3a in Table~\ref{tab:uncertainty}), we obtained a position uncertainty of 0.039 mas, corresponding to 12\% of the beam size. This is close to the uncertainties obtained for isolated bright and compact components in previous studies \citep[e.g.,][]{lister09}. In contrast, the position uncertainty for the time range between 2017 November and 2022 June (denoted as C3b in Table~\ref{tab:uncertainty}) is 0.180 mas, which corresponds to 56\% of the beam size. This is because C3 is bright and exhibits a relatively simple structure during the former period, whereas it becomes fainter and its structure becomes more complex during the latter period \citep[see light curves in][]{hodgson18, kino18, kino21}.

The position uncertainties of K1--K4 are different for different subcomponents. The uncertainties of K1 are smaller than 20\% of the beam size regardless of the methods used, while the uncertainties of K2--K4 are larger than 20\% of the beam size. Arguably, this is because K1 was relatively isolated and ejected directly through the ambient medium, whereas K2--K4 were formed late and propagated through more diffused media. It is also possible that the position uncertainty of K1 is underestimated (see Appendix~\ref{appendix_c}). The small number of data points for K4, combined with the complex morphology of the media it passes through, likely contributed to the largest position uncertainty among K1--K4.

\begin{figure*}[t!]
\centering
\includegraphics[width=0.8\textwidth]{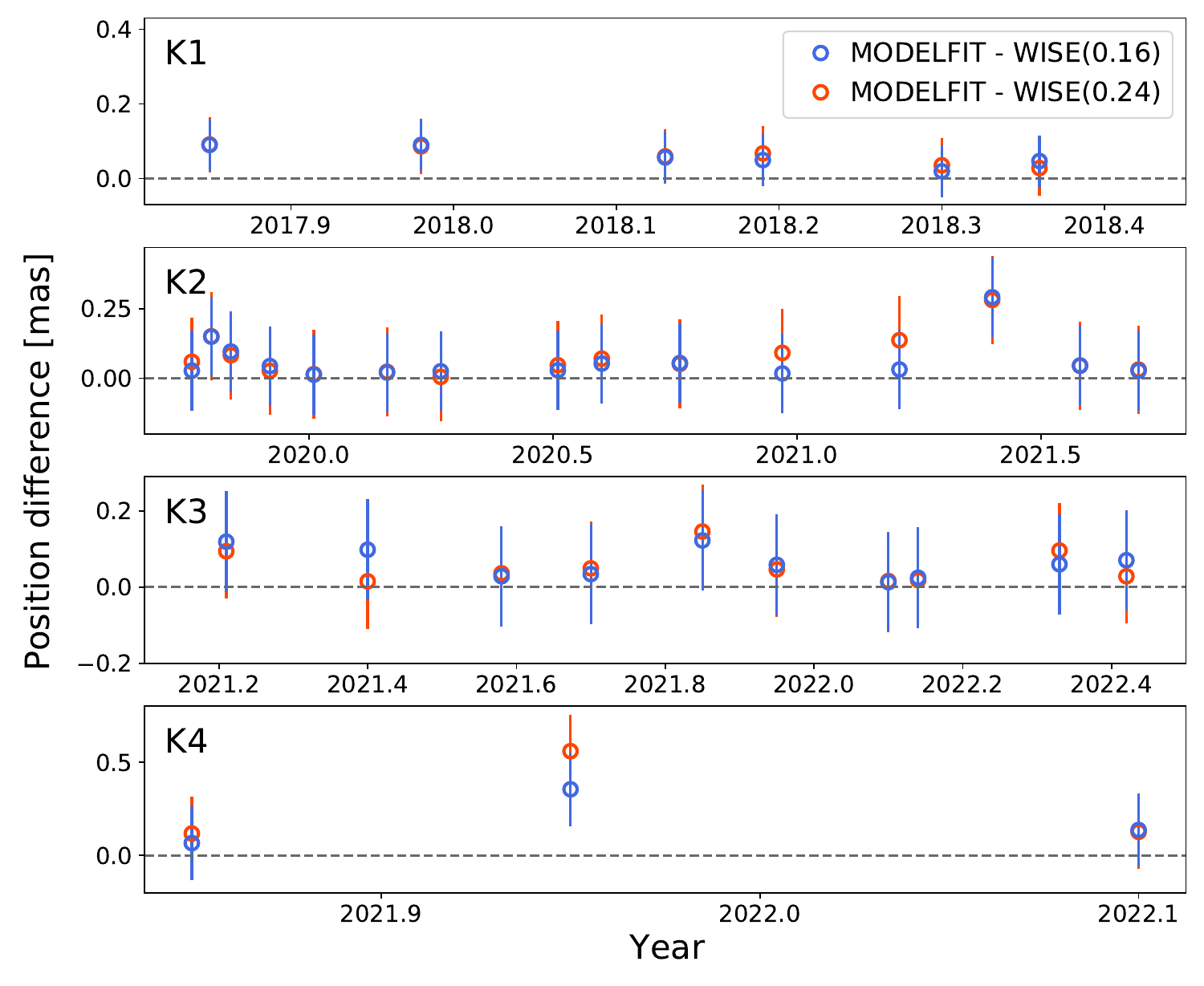}
\caption{Absolute differences between the positions of K1--K4 obtained with MODELFIT and WISE at 0.16 and 0.24 mas, represented by blue and red circles, respectively. The error bars represent the square root of the sum of the squares of the position uncertainties obtained with MODELFIT and WISE. The gray dashed lines indicate a difference of zero. We find that most of the position differences are smaller than the error bars.
\label{appendix:common_difference}}
\end{figure*}

To quantify potential systematic differences between the three methods used for position measurements of K1--K4, we compare their results for every epoch in which all three algorithms identify a given component; the results are summarized in Figure~\ref{appendix:common_difference}.
We find that the position values agree within their formal uncertainties derived from \citet{homan01}.

\section{Apparent velocity}\label{appendix_c} 

The differences in the apparent speeds of K1--K4 derived from MODELFIT and WISE are larger than the measurement uncertainties (Table~\ref{tab:summary}). To investigate this, we compare the jet features obtained for the individual time period with those obtained for the time periods where all MODELFIT and WISE at scales of 0.16 and 0.24 mas detected the same component in Table~\ref{tab:velocity}. The apparent speeds of K2-K4 measured for the common period are consistent within the uncertainties. This suggests that the difference in their apparent speeds indeed arises from the difference in the number of epochs for which each subcomponent is identified. In the case of K1, however, the apparent speeds measured with WISE are larger than that measured with MODELFIT by more than the uncertainties. This suggests that the actual position uncertainty of K1 would be larger than that derived from the method described in \citet{homan01}. Nevertheless, even in this case, both methods reach the same conclusion that K1 is slower than K2-K4. This consistency reaffirms the reliability of the main results. 

\begin{table*}[t!]
    \centering 
    \caption{Jet features for different methods and periods\label{tab:velocity}}
    \begin{tabular}{cccccccc}
        \hline
        Method & Name & Individual period & $\beta_{\rm app}$ & P.A. ($^\circ$) & Common period & $\beta_{\rm app}$ & P.A. ($^\circ$) \\
        \hline 
        \multirow{4}{*}{MODELFIT} & K1 & 2017.85 - 2020.16 & $0.46\pm0.02$ & $185.5\pm2.2$ & 2017.85 - 2018.36 & $0.66\pm0.18$ & $141.0\pm11.2$ \\
        & K2 & 2019.76 - 2022.14 & $0.91\pm0.06$ & $155.9\pm2.5$ & 2019.76 - 2021.70 & $1.00\pm0.07$ & $157.5\pm3.0$ \\
        & K3 & 2021.21 - 2022.64 & $1.35\pm0.14$ & $164.4\pm4.3$ & 2021.21 - 2022.42 & $1.35\pm0.14$ & $164.4\pm4.3$ \\
        & K4 & 2021.85 - 2022.14 & $1.99\pm1.09$ & $182.9\pm21.9$ & 2021.85 - 2022.10 & $2.38\pm1.85$ & $184.1\pm31.5$ \\
        \hline 
        \multirow{4}{*}{WISE (0.16)} & K1 & 2017.43 - 2019.80 & $0.59\pm0.03$ & $171.2\pm2.0$ & 2017.85 - 2018.36 & $0.91\pm0.20$ & $144.4\pm5.7$  \\ 
        & K2 & 2019.50 - 2021.70 & $1.05\pm0.06$ & $158.3\pm2.2$ & 2019.76 - 2021.70 & $1.02\pm0.07$ & $156.2\pm1.1$  \\
        & K3 & 2020.42 - 2022.42 & $1.11\pm0.06$ & $168.6\pm2.1$ & 2021.21 - 2022.42 & $1.31\pm0.11$ & $166.8\pm2.2$ \\
        & K4 & 2020.76 - 2022.10 & $1.22\pm0.14$ & $167.7\pm4.7$ & 2021.85 - 2022.10 & $2.43\pm1.00$ & $168.2\pm35.2$ \\
        \hline
        \multirow{4}{*}{WISE (0.24)} & K1 & 2017.43 - 2018.36 & $0.82\pm0.07$ & $145.9\pm3.5$ &  2017.85 - 2018.36 & $0.90\pm0.14$ & $147.2\pm6.3$  \\ 
        & K2 & 2019.76 - 2021.70 & $1.02\pm0.06$ & $156.9\pm2.2$ & 2019.76 - 2021.70 & $1.02\pm0.03$ & $156.9\pm1.1$ \\
        & K3 & 2020.27 - 2022.42 & $1.08\pm0.05$ & $168.3\pm2.1$ & 2021.21 - 2022.42 & $1.32\pm0.07$ & $166.0\pm2.2$ \\
        & K4 & 2020.76 - 2022.14 & $1.27\pm0.13$ & $168.7\pm4.0$ & 2021.85 - 2022.10 & $2.26\pm0.52$ & $173.1\pm9.1$ \\
        \hline
    \end{tabular}
\tablecomments{Jet features of each component identified with MODELFIT and WISE at scales of 0.16 and 0.24 mas. Individual period refers to the entire period for which each component is identified with each method, while common period refers to the period for which each component is consistently identified with different methods.} 
\end{table*}

\end{document}